**Microwave resonator-enabled broadband on-chip electro-optic frequency comb generation**

Zhaoxi Chen[1,†,*], Yiwen Zhang[1,†], Hanke Feng[1], Yuansong Zeng[1], Ke Zhang[1], and Cheng Wang[1,2,*]

[1]Department of Electrical Engineering, City University of Hong Kong, Kowloon, Hong Kong, China

[2]State Key Laboratory of Terahertz and Millimeter Waves, City University of Hong Kong, Kowloon, Hong Kong, China

*zxchen4@cityu.edu.hk

*cwang257@cityu.edu.hk

† These authors contributed equally to this work.

**Abstract**

Optical frequency combs play a crucial role in optical communications, time-frequency metrology, precise ranging, and sensing. Among various generation schemes, resonant electro-optic combs are particularly attractive for its excellent stability, flexibility and broad bandwidths. In this approach, an optical pump undergoes multiple electro-optic modulation processes in a high-$Q$ optical resonator, resulting in cascaded spectral sidebands. However, most resonant electro-optic combs to date make use of lumped-capacitor electrodes with relatively inefficient utilization of the input electrical power. This design also reflects most electrical power back to the driving circuits and necessitates costly RF isolators in between, presenting substantial challenges in practical applications. To address these issues, we present an RF circuit friendly electro-optic frequency comb generator incorporated with on-chip coplanar microwave resonator electrodes, based on a thin-film lithium niobate platform. Our design achieves more than three times electrical power reduction with minimal reflection at the designed comb repetition rate of ~ 25 GHz. We experimentally demonstrate broadband electro-optic frequency comb generation with a comb span of >85 nm at a moderate electrical driving power of 740 mW (28.7 dBm). Our power-efficient and isolator-free electro-optic comb source could offer compact, low-cost and simple-to-design solution for applications in spectroscopy, high-precise metrology, optical communications.

**Introduction**

Optical frequency comb (OFC) generators play crucial roles in various applications, including optical communications [1-3], spectroscopy [4, 5], timekeeping [6], precise ranging [7, 8], and exoplanet detections [9], by providing excellent light sources with coherent and equally spaced spectral lines. Recent progress in photonic integrated circuits has paved the way for on-chip comb sources, with significantly improved compactness, efficiency, and scalability [10-13]. Integrated OFCs have been achieved based on various physical principles and material platforms, including mode-locked semiconductor lasers (e.g. in GaAs [14], and InAs [15]), optical Kerr nonlinearity (e.g. in SiN [16, 17], AlN [18], and AlGaAs [19]), and electro-optic (EO) modulation (e.g. in LN [20-23] and Si [24] ).

Among these schemes, electro-optic frequency combs are particularly attractive for its GHz repetition rates, broad tunability, and intrinsic mutual coherence [25, 26]. An EO frequency comb is generated by modulating a continuous-wave laser signal through one or multiple phase and amplitude EO modulators. This modulation process translates the input laser's single frequency into a comb of equally spaced frequency lines. Traditionally, this is often achieved using off-the-shelf modulators based on lithium niobate (LiNbO$_3$, LN), a material well known for its excellent optical properties and significant $\chi^{(2)}$ nonlinearity [27, 28]. In recent years, the rapidly emerging thin-film LN (TFLN) platform, with tightly confined optical waveguides and substantially enhanced EO modulation efficiency [29-33], has further enabled integrated EO combs with much higher integration level and wider comb span compared with their bulk counterparts [34, 35]. In particular, resonant EO combs that leverage optical resonators with ultrahigh quality ($Q$) factors have achieved remarkable comb span, since the optical pump can be circulated and modulated for dozens of round trips. For example, over 80 nm wide EO comb has been achieved in a TFLN optical resonator with a $Q$ factor of 1.5×10$^6$ [22]. Further adopting a dual-resonator design has pushed the optical conversion efficiency to 30% with an even wider comb span of 132 nm in wavelength [36].

To date, however, most on-chip resonant EO comb generators make use of a ground-signal-ground (GSG) capacitive electrode for applying the EO modulation signals (Fig. 1a). This electrode configuration is essentially a lumped-capacitor load from the driving circuit perspective, where the input electrical power is almost fully reflected and not efficiently utilized. As a result, several watts of electrical driving power are often required for broadband EO comb generation (e.g. 2.2 W in Ref. [36] and 1 W in Ref. [37]). Moreover, the high reflected electrical power could be detrimental to the driving RF circuit, necessitating bulky and costly isolators or circulators to prevent power reflection to the electrical amplifier [Fig. 1(a)]. In short, the lumped capacitor electrode design has become a major hurdle in terms of complexity, cost, and power consumption to the practical application of integrated resonant EO combs.

To address these issues, we propose and demonstrate an on-chip coplanar waveguide (CPW) microwave resonator electrode for efficient and RF-circuit friendly driving of EO frequency combs. Compared with a conventional lumped-capacitor electrode, our device features a 3.6 times electrical field enhancement, which translates into more than 3 times reduction in power consumption with negligible electrical power reflection (-50 dB). Leveraging a 4-inch wafer-scale TFLN platform, we experimentally demonstrate broadband power-efficient EO frequency comb generation with a repetition rate of 25.6 GHz and a comb span exceeding 85 nm. Importantly, this is achieved using an optical racetrack resonator with a moderate $Q_L$ = 8.5 ×10$^5$, at a relatively low electrical driving power of 28.7 dBm, and without the use of electrical isolators or circulators. The design and analytical model can be readily extended to other frequencies, supporting power-efficient EO comb generation with a wide range of target repetition rates.

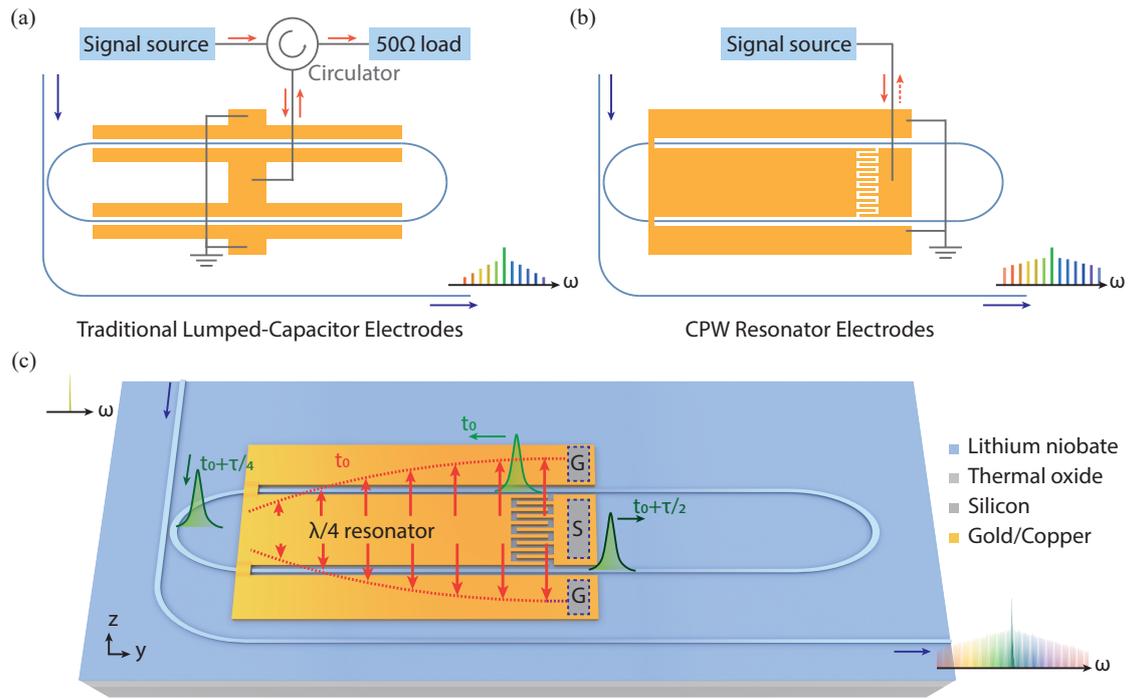

**Fig.1 Working principle of the microwave resonator-enabled broadband EO comb generation process.** (a-b) Schematic comparison of EO comb generators with traditional lumped capacitor design (a), where a circulator and an external 50-Ω load are needed to prevent electrical power reflection to the driving circuit, and the proposed $\lambda/4$ microwave resonator design (b). The microwave resonator serves both electrical field enhancement and prevention of electrical power reflection. (c) Schematic illustration of the electrical field distribution and phase-matching condition in the microwave resonator enhanced EO comb generator. Green pulses indicate the optical signal locations at different time $t_0$, $t_0+\tau/4$, and $t_0+\tau/2$. Red arrows indicate electrical field distributions in the two gaps of the coplanar microwave resonator at $t_0$, which would be reversed at $t_0+\tau/2$.

**Design of coplanar waveguide resonator electrode**

Figure 1c illustrates the working principle of the proposed EO comb generator, consisting of an optical racetrack resonator fabricated in TFLN, integrated with CPW resonator electrodes. High-speed modulating electrical signal is applied to the input GSG microelectrodes, which is capacitively coupled into the coplanar waveguide resonator by an interdigitated electrode (IDE). The coplanar waveguide resonator consists of a GSG transmission line with a length equal to a quarter wavelength ($\lambda/4$) and shorted terminals on the opposite end, forming a quarter-wave resonator. When the applied microwave signal is near resonance frequency, the electrical field is significantly enhanced at the coupling (open) port, while the shorted end exhibits zero voltage (but finite current flow), following a sinusoidal $\lambda/4$ standing-wave pattern along the transmission line (Fig. 1c). Through precise engineering of the IDE to achieve impedance matching between the microwave resonator and the driving circuit, the input electrical power can be critically coupled into the resonator, leading to enhanced EO modulation efficiency and negligible reflected electrical power.

A crucial requirement to achieve efficient EO comb generation is phase matching between the resonant optical- and micro-waves. In our device architecture, this is naturally satisfied when the microwave resonance frequency $f_{MR}$, applied microwave frequency $f_{MW}$, and the optical free-spectral range (FSR) are equal to each other. As the schematic shows in Fig. 1c, considering a counter-clockwise-traveling optical pulse (green) present at the top middle section of the microwave resonator at time $t_0$, it experiences a positive maximum EO modulation effect if the microwave field is pointing upwards (from signal to ground) at this point. The optical pulse circulates and reaches the bottom middle section of the resonator at time $t_0+\tau/2$, where $\tau=1/FSR$ is the round-trip time of the optical resonator. Although the electric field at this location is opposite to that in the upper gap (pointing downwards at time $t_0$, as shown in Fig. 1c), it exactly flips to the upward-pointing direction at $t_0+\tau/2$, as long as the microwave modulation frequency $f_{MW}$ is equal to the optical FSR (such that microwave period $T_{MW}=1/f_{MW}=\tau$). As a result, the optical pulse again sees a positive maximum EO modulation field at $t_0+\tau/2$. At other locations of the resonator, although the electric field strength may be smaller, the optical signal always experiences upward-pointing electric field and therefore a constructive accumulation of EO modulation throughout the resonator.

In our device targeting a repetition rate of ~25 GHz, the microwave transmission line has an effective index ($n_{eff,MW}$) of ~2.6 and an effective wavelength of 4.4 mm, leading to a total length of 1.1 mm for the $\lambda/4$ resonator. Meanwhile, the optical group index ($n_{g,O}$) is 2.26 in the TFLN waveguide at telecommunication wavelengths, necessitating a round-trip length of 5.1 mm for the racetrack resonator. We use a bending radius of 80 micrometer with a Euler curve shape to minimize bending loss, such that the straight section of the racetrack is 2.3 mm long. This allows the $\lambda/4$ resonator to be placed approximately within the left half of the optical resonator to satisfy the phase-matching condition discussed above.

We design and fine tune the microwave resonator and coupling IDE to achieve impedance matching with the external driving circuit at the target resonance frequency. Figure 2a illustrates the top-view and cross-section schematic of the CPW microwave resonator. This short-circuit $\lambda/4$ resonator can be equivalently modeled by a parallel $RLC$ resonant circuit near resonance (Fig. 2b), where the input signal is applied from the driving circuit (ii) into the resonator circuit (i). Assuming there is no extra loss on the transmission line, the total input impedance of (i) is given by

$$Z_{in} = \frac{1}{i\omega C_\kappa} + Z_{LCR} = \frac{1}{i\omega C_\kappa} + \left(\frac{1}{i\omega L} + i\omega C + \frac{1}{R}\right)^{-1} \quad (1)$$

where $C_\kappa$ represents the capacitance of the IDE, and $R$, $L$ and $C$ are the equivalent resistance, inductance and capacitance of the $RLC$ resonator, respectively. Details of the derivation of characteristic parameters are discussed in the Supplementary Information. It should be noted that the existence of the coupling capacitor not only changes the input impedance $Z_{in}$, but also shifts the resonance frequency $f_{MR}$ from the isolated $RLC$ resonance. As a result, the on-resonance impedance $Z_{in}$ could be effectively controlled by fine tuning the coupling capacitance $C_\kappa$. In our experiments, we vary the IDE length $L_f$ to achieve a near 50 Ω input impedance to match that of the external driving circuit and minimize power reflection.

We experimentally verify the performance of the designed $\lambda/4$ resonator by measuring the reflection coefficient $S_{11} = |(Z_{in} - R_L)/(Z_{in} + R_L)|$, which should ideally be zero at 25 GHz, using a vector network analyzer. We vary the coupling strength by sweeping the finger length $L_f$ of the IDE with fixed finger width $w_f=3$ μm and slot width $s_f=g_f=2$ μm. Figure 2c displays the measured responses from a CPW resonator with a finger length of 33.5 μm (blue) and a lumped capacitor electrode (red). The fabricated resonator shows a strong resonance dip at 25.67 GHz, with an on-resonance reflection down to -49.8 dB. The measured results are also consistent with the calculation results from the equivalent circuit model (yellow), where the slight discrepancy may result from deviations in the geometric dimensions and dielectric constants between theory and actually fabricated devices. Importantly, the power reflection could remain < -20 dB (less than 1%) within a relatively broad frequency range of 1 GHz, which provides crucial tolerance and flexibility in practical applications where the optical FSR may not be perfectly aligned with the microwave resonance. In our actual device to be discussed in more details next, the EO comb operates at a repetition rate of 25.612 GHz (dashed line), where the power reflection is -46 dB. This is in sharp contrast to the lumped-capacitor case (red) with a -3 dB power reflection into the driving circuit (rest is lost in the on-chip resistance).

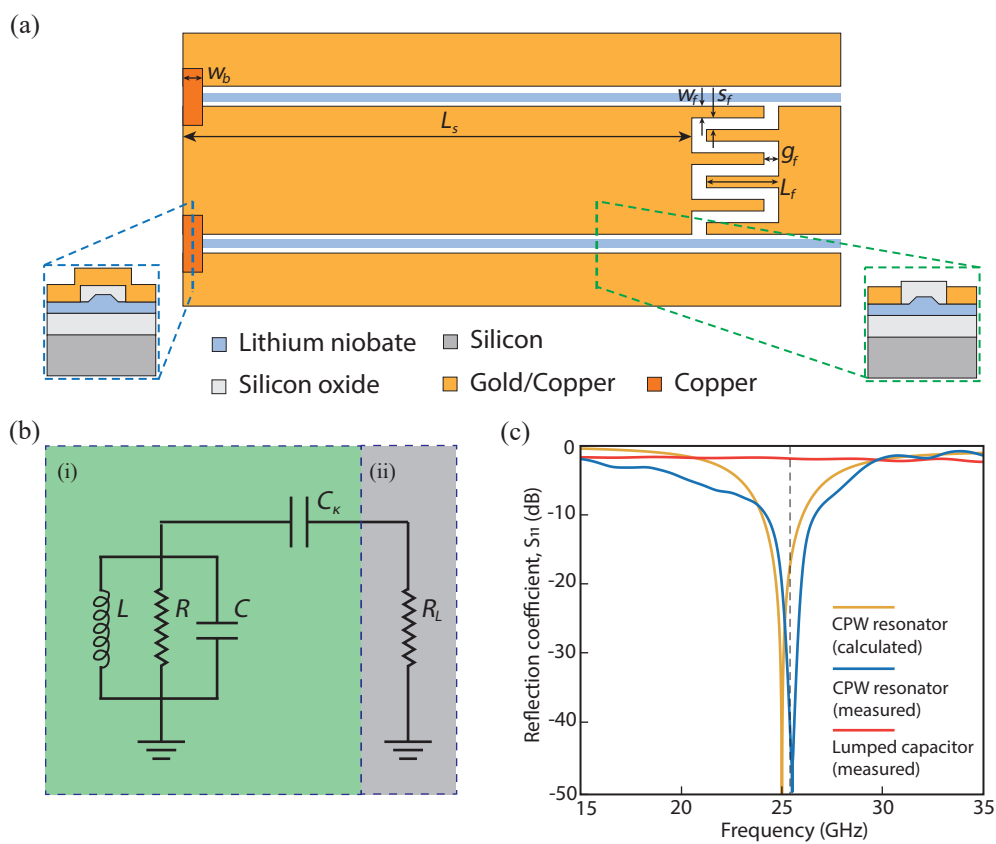

**Fig. 2 Microwave resonator design.** (a) Top view of the CPW resonator with shorted terminal at the left end and coupling interdigitated finger electrodes at the right-hand side. Insets: Cross-section schematics of the CPW resonator at the shorted metallic bridge (blue dashed line) and along the transmission line (green dashed line), respectively. (b) Equivalent circuit model of the short-circuited λ/4 CPW resonator (i), driven by external circuit (ii) with a source impedance of $R_L$ = 50 Ω. (c) Calculated (yellow) and measured microwave reflection coefficients of the CPW resonator (blue), and that measured from a lumped capacitor electrodes (red), respectively.

**Device fabrication and electro-optic frequency comb generation**

All devices are fabricated on x-cut TFLN wafers (NANOLN). The wafer stack consists of a 500-nm TFLN layer, a 4.7-μm thermal oxide buffer layer and a 500-μm high-resistance silicon substrate layer. The bare wafer was first coated by a layer of 700 nm thick $SiO_2$ using plasma-enhanced chemical vapor deposition (PECVD) as etch mask. The optical waveguides and optical racetrack cavities are then patterned by an ASML UV Stepper lithography system (NFF, HKUST). The patterns are transferred into the oxide mask layer and LN layer sequentially using reactive ion etching (RIE) with a 250 nm etch depth. After removing the remaining etch mask, another layer of PECVD oxide is coated to form a 1.5 μm thick upper cladding of the optical waveguides. The metallic electrodes (750 nm of copper, 50 nm of gold) are formed by a second stepper lithography process, followed by thermal deposition and lift-off. The signal has a width of 150 μm and the gap between signal and ground is set as 7 μm. 5 μm wide metallic bridges (800 nm of copper) are then patterned at the short end of the CPW resonator by electron-beam lithography (EBL), thermal deposition and lift-off processes. Finally, facets of the fabricated devices are cleaved for optical measurements. The fabricated optical bus (racetrack) waveguide has a top width of 1.2 μm (2 μm) and the racetrack bends are designed with Euler-curve shape to reduce radiation loss. Figure 3a shows the micrographs of the fabricated EO comb generators with lumped-capacitor (top, as reference) and CPW resonator electrodes (bottom), respectively. Figure 3b shows scanning electric microscope (SEM) images with details of the metallic bridges and the IDE.

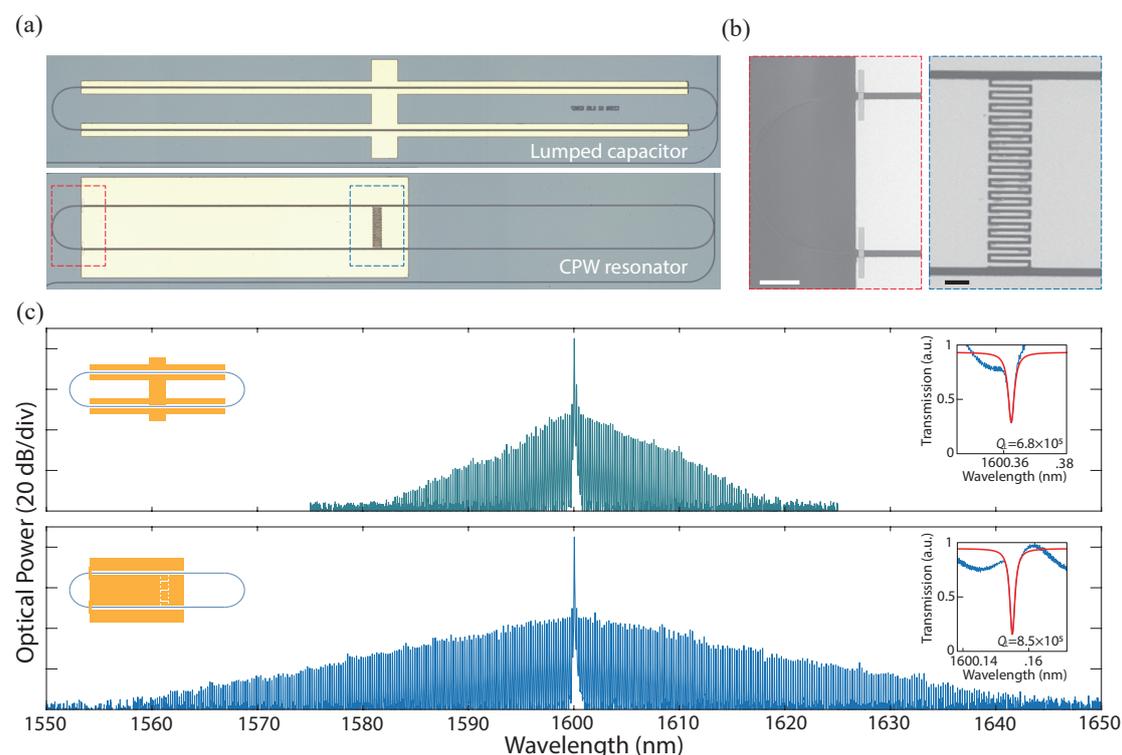

**Fig. 3 Broadband EO comb generation.** (a) Micrographs of the fabricated on-chip EO comb generators with lumped-capacitor (top) and CPW resonator (bottom) electrodes. (b) SEM images of the metallic bridges at the shorted terminal (left) and the IDE coupler (right), respectively. The scale bars are 20 μm in both panels. (c) Measured EO comb spectra from the lumped capacitor (top) and CPW resonator electrodes (bottom) with the same input electrical power of 28.7 dBm. Insets show the device configurations (left) and measured optical transmission spectra (right) of the corresponding devices.

We finally demonstrate power-efficient and RF isolator-free EO comb generation enabled by the CPW microwave resonator. Optical pump from a continuous-wave tunable laser (SANTEC-550) is launched into the TFLN waveguide via a lensed fiber. The optical output is collected by another lensed fiber and recorded by an optical spectrum analyzer (Yokogawa AQ6370D). The driving microwave signal near 25 GHz is generated by a radio-frequency synthesizer and amplified by an electrical power amplifier. For the microwave-resonator devices, the amplified electrical signal is applied directly onto the microelectrodes using a GSG probe, without the need of a circulator or isolator, thanks to the excellent impedance matching and minimal power reflection. As for the reference device with lumped capacitor electrodes, the amplified electrical signal is first passed through a microwave circulator before delivered to the microelectrodes. The reflected electrical power is then terminated at a 50-Ω load through the circulator, similar to that used in Refs. [22, 36]. To compensate the loss from the circulator (0.9 dB at 25 GHz), the output power from electrical amplifier is slightly higher than that used for measuring the CPW devices. During the experiments, the optical input pump power is fixed at 2 mW and the applied electrical driving power on chip are calibrated to be 740 mW (28.7 dBm) in both cases.

We show that our CPW resonator enables frequency comb generation with approximately doubled comb span from the reference lump-capacitor device. When the input optical and microwave frequencies are both tuned into resonance with the on-chip optical and microwave resonators, we achieve broadband EO comb with an 85 nm span and 430 comb lines at a repetition rate of 25.612 GHz (Fig. 3c, bottom). The measured spectral span and roll-off slope (~0.9 dB/nm) are both on par with that reported in Ref.[22], but achieved using similar input RF power and an optical resonator with 1.8 times lower loaded $Q_L$ factor ($Q_L$ = 8.5×$10^5$). Since the resonant EO comb span is proportionate to both optical $Q_L$ and modulation index [38], this indicates an approximately 1.7 times increased modulation index. The modulation enhancement factor is further corroborated using the measured EO comb from the reference device fabricated on the same chip, which features a comb span of 38 nm (narrower by a factor of 2.2) with a repetition rate of 25.255 GHz (Fig. 3c, top). Taking into account the slightly lower $Q_L$ of 6.8×$10^5$ in the reference device, the effective modulation enhancement factor of our microwave resonator design is 1.8, which corresponds to more than 3 times reduction in electrical power consumption. Considering the signal length of the microwave resonator electrode is approximately half (1100 μm) of that in the lumped capacitor device (2300 μm), we estimate that the average electric field strength in the EO modulation region is enhanced by a factor of 3.6 in the microwave resonator. The EO comb span could in principle be further broadened to 240 nm with an improved optical $Q$ factor of 1.4×$10^6$ and a higher microwave power of 33.4 dBm (same condition as that in Ref.[36]). In this case, a more careful engineering of the waveguide dispersion will likely be required. We note though, operating in a moderate-$Q$-factor regime in this work also offers distinct advantages for practical applications, as the overall pump-to-comb conversion efficiency is higher (~0.6%, as compared to 0.3% in Ref. [22]) and the system is less prone to optical and microwave detuning.

Theoretically, our EO comb generator exhibits optimal performance when the microwave resonance frequency $f_{MR}$ is perfectly matched with the optical FSR. To investigate the tolerance to potential mismatch between $f_{MR}$ and FSR that may arise from fabrication deviations, we fabricate and measure a series of devices with identical microwave resonator electrodes, but varying optical racetrack resonator sizes. The performance of each EO comb generator is evaluated by fixing the optical pump and microwave driving power, while fine-tuning microwave frequency to the optical FSR in each device. As shown in Fig. 4, the generated comb span gradually decreases when the optical FSR shifts away from the CPW resonance frequency at 25.67 GHz. The trend of measured comb span (blue dots) agrees well with the calculated relative electric field strength (black curve). Importantly, the measured comb span remains >75 nm (~90% the optimal value) within a relatively broad optical FSR range of 25.4-25.7 GHz, providing important robustness and tolerance in practical scenarios.

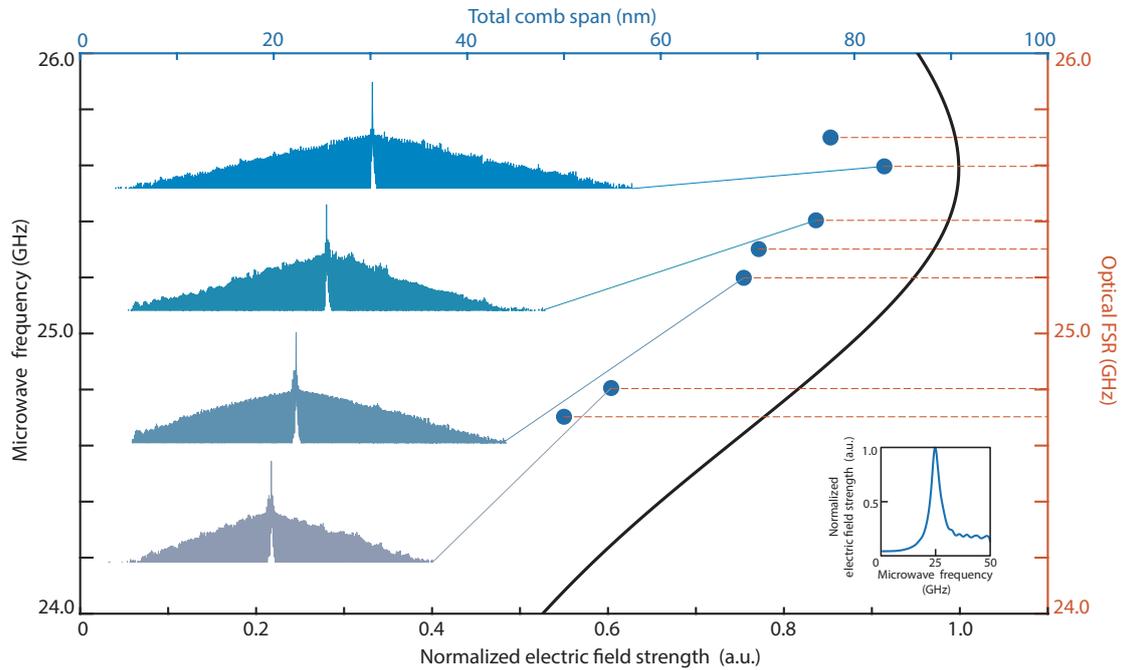

**Fig. 4 Tolerance to frequency mismatch between optical and microwave resonators.** Blue dots show the measured comb span (top horizontal axis) for optical racetrack resonators with different FSRs (right vertical axis), driven with microwave frequencies (left vertical axis) matched with FSR. The corresponding comb spectra are shown in the left insets. Black solid curve shows the normalized electric field strength (bottom horizontal axis) derived from the measured $S_{11}$ parameter of the CPW resonator (results in the full measured frequency range are shown in the bottom right inset).

**Discussions and conclusions**

In conclusion, our EO comb source embedded with $\lambda/4$ CPW microwave resonator electrodes not only feature substantially improved comb generation efficiency, but also strongly suppresses power reflection back to the driving circuit. The reduced RF power consumption and removal of electrical isolators or circulators are highly desired for such EO combs to be applied in practical and commercial scenarios. The design principles and analytical model developed in this work can be readily extended to other microwave frequencies (and in turn comb repetition rates) by scaling the CPW resonator and IDE lengths. Our research provides a simple and feasible solution for achieving efficient and low-cost integrated EO frequency combs, promising for applications in precision metrology, advanced spectroscopy, and optical communication systems.

**Acknowledgements**

This work is supported in part by the Research Grants Council, University Grants Committee (CityU 11212721, CityU 11204022, N_CityU113/20); Croucher Foundation (9509005); City University of Hong Kong (9610682). We acknowledge Nanosystem Fabrication Facility (CWB) of HKUST for the device/system fabrication.

**Conflict of interest**

The authors declare no conflicts of interest.

**Data availability**

Data underlying the results presented in this paper are not publicly available at this time but may be obtained from the authors upon reasonable request.